\documentclass[aps,preprint,showpacs,preprintnumbers,amsmath,amssymb]{revtex4}
\usepackage{amsmath,mathrsfs,amsbsy,color,graphicx,bm,amsthm,amsfonts}
\usepackage{units}
\usepackage{bbm}
\usepackage{times}
\usepackage{dcolumn}
\usepackage{mathrsfs}
\usepackage{amsmath,amssymb,epsfig}
\usepackage{amsmath}
\newcommand{\udots}{\mathinner{\mskip1mu\raise1pt\vbox{\kern7pt\hbox{.}}
\mskip2mu\raise4pt\hbox{.}\mskip2mu\raise7pt\hbox{.}\mskip1mu}}

\begin{document}

\title{Gaussian quantum steering in multi-event horizon spacetime
}
\author{Shu-Min Wu$^1$\footnote{smwu@lnnu.edu.cn}, Jin-Xuan Li$^1$, Xiao-Wei Fan$^1$, Wen-Mei Li$^1$,  Xiao-Li Huang$^1$\footnote{huangxiaoli1982@foxmail.com}, Hao-Sheng Zeng$^2$\footnote{Corresponding author:hszeng@hunnu.edu.cn}}
\affiliation{$^1$ Department of Physics, Liaoning Normal University, Dalian 116029, China\\
$^2$ Department of Physics, Hunan Normal University, Changsha 410081, China}


\begin{abstract}
We study Gaussian quantum steering in the Schwarzschild-de Sitter (SdS)  spacetime that is endowed with both a black hole event horizon (BEH) and a cosmological event horizon (CEH), giving rise to two different Hawking temperatures. It is shown that the Hawking effect of the black hole always reduces the quantum steering, but the Hawking effect of the expanding universe does not always play the same role. For the first time, we find that  the Hawking effect can improve quantum steering. We also find that the observer who locates in the BEH has stronger steerability than the observer who locates in CEH. Further, we study the steering asymmetry, and the conditions for two-way, one-way and no-way steering in the SdS spacetime.
Finally, we study the Gaussian quantum steering in the scenario of effective equilibrium temperature. We show that quantum steering reduces monotonically with the effective temperature but now increases monotonically with the Hawking temperature of  the black hole, which banishes the belief that the  Hawking effect can only destroy quantum steering.
\end{abstract}

\vspace*{0.5cm}
\maketitle
\section{Introduction}
Schr\"{o}dinger first introduced Einstein-Podolsky-Rosen (EPR) steering  to argue the action at a distance paradox in the famous work by Einstein, Podolsky, and Rosen \cite{L1,L2}.
EPR steering, a category of the nonlocal correlations, representing the ability of one observer to affect another observer's state via local measurements, is different from both Bell nonlocality and quantum entanglement by possessing an asymmetric property.
Like quantum entanglement, EPR steering is also a very important quantum resource that can be used for quantum information tasks, such as subchannel discrimination and one-sided device-independent quantum key distribution \cite{L4,L5,L6,L7,L8,L9}.
Despite the fundamental importance of quantum steering, it received increasing attention until
Wiseman $et$ $al$. developed this concept in 2007 and introduced a rigorous operational definition \cite{L10}. Subsequently, compared with quantum entanglement, it was found that quantum steering has richer properties. On account of  its unique directional property, quantum steering shows an asymmetric manifestation that further leads to one-way steering, that is Bob can steer Alice but not vice versa \cite{L11,L12,L13,L14,L15}.

In the last years, people have been very much interested in relativistic quantum information \cite{Q38,Q39,Q40,QWE82,Q41,Q42,Q43,Q50,Q48,Q44,Q45,Q46,Q47,Q49,Q51,Q52,Q53,Q54,Q55,Q56,Q57,Q58,wsm1,wsm2,wsm3,wsm4,wsm5,wsm6,wsm7,
Q59,Q60,Q61,Q63,Q64,Q65}. Especially, the influences of the Hawking and the Unruh effects on quantum correlation and coherence have been widely studied \cite{Q38,Q39,Q40,QWE82,Q41,Q42,Q43,Q50,Q48,Q44,Q45,Q46,Q47,Q49,Q51,Q52,Q53,Q54,Q55,Q56,Q57,Q58,wsm1,wsm2,wsm3,wsm4,wsm5,wsm6,wsm7}. It has been shown that the Hawking effect destroys quantum steering between  Alice at an asymptotically flat region and Bob who hovers outside the event horizon, in which bosonic steering and fermionic steering have different behaviors: The former asymptotically vanishes \cite{AQQWE70}, while the latter can survive forever in Schwarzschild spacetime \cite{Q47}. It also has been shown that the Unruh effect has a similar action on quantum steering \cite{AQQWE71}, and the quantum steering for two relatively accelerated detectors in flat space has been studied \cite{AQQWE72}. Note that all these researches are restricted to the spacetime of the single-event horizon.

The de Sitter solution is a simplest solution of general relativity field equations  with a nonzero cosmological constant \cite{L16,L17,L18,L19}. Since our universe is undergoing an accelerated expansion \cite{L20,L21}, the study of phenomena in spacetime asymptotically de Sitter is more realistic than that in spacetime asymptotically flat. Therefore, the Schwarzschild-de Sitter (SdS) spacetime, which is described by the cosmological constant and the geometric mass of the central Schwarzschild black hole \cite{L22,L23,L24,L25}, is of great interest.
In this more realistic scenario, the solutions of the black hole  are asymptotically de Sitter, rather than asymptotically flat. The SdS spacetime
is the existence of the black hole event horizon (BEH) and the cosmological event horizon (CEH), which admit two-temperature thermodynamics qualitatively much different from the single horizon. Compared to the cases of single-event horizon spacetime, quantum information processing in multi-event horizon spacetime may have richer properties.

In this work, we present a quantitative investigation for Gaussian quantum steerability of free bosonic fields in the Schwarzschild-de Sitter (SdS) spacetime  endowed with both a BEH and a CEH \cite{QWE70}. We initially consider a two-mode squeezed Gaussian state with squeezing  parameter $s$ shared by Alice and Bob \cite{Q40}.  Our model involves two modes: the mode $A$ observed by Alice located at the BEH; the mode $B$ observed by Bob located at the CEH.
 A Kruskal vacuum state observed by a Kruskal observer would be detected as a
thermal state from Alice and Bob's viewpoint. Such a process in quantum information scenario  can be described as the Gaussian channels acting on quantum state shared by Alice and Bob \cite{QWE71,QWE72}. We will study Gaussian quantum steering from two independent descriptions of thermodynamics and particle creation in this background. The first involves
thermal equilibrium of an observer with either of the horizons, and the second treats both the horizons combined so as to define an
effective equilibrium temperature. We find that the two descriptions lead to very different results, which suggests the fact that quantum information is observer dependent.

The paper is organized as follows. In Sec. II, we
briefly introduce the measure of Gaussian quantum steering. In Sec. III, we discuss how
the gravitational effect of the  Schwarzschild-de Sitter black hole spacetime can be
described by the Gaussian  channels.
In Sec. IV, we study the behaviors of Gaussian quantum steering and its
asymmetry in multi-event horizon spacetime.  In Sec. V, we study Gaussian quantum steering in the scenario of the effective temperature.
The last section is devoted to a brief conclusion.
\section{Measure of Gaussian quantum steering \label{GSCDGE}}
We consider a two-mode Gaussian quantum system of continuous variables that supports on the Hilbert space $\mathcal{H}=\mathcal{H}_1\otimes\mathcal{H}_2$, where $\mathcal{H}_i$ is the infinite dimensional Hilbert space of each bosonic subsystem \cite{QWE73}. Quadrature operators for each mode are described by  $\hat q_i=\hat a_i+\hat a^{\dagger}_i$ and $\hat p_i=-i(\hat a_i-\hat a^{\dagger}_i)$, where $\hat a_{i}$ and  $\hat a^{\dagger}_i$ are respectively the annihilation and creation operators of each modes, which meet the commutation relations $[\hat a_i, \hat a^{\dagger}_j]=\delta_{ij}$. These quadratures can be grouped in the vector of operators
$\hat R = (\hat q_1,\hat p_1, \hat q_2,\hat p_2)$, and the canonical commutation relation then becomes as  $[{{{\hat R}_i},{{\hat R}_j}} ] = 2i{\Omega _{ij}}$, with $\Omega = {{\ 0\ \ 1}\choose{-1\ 0}}^{\bigoplus{2}}$ being the symplectic matrix.
The characteristics of a two-mode Gaussian state $\rho _{AB}$ can be fully described by the expectation $\langle\hat R\rangle$ and the second statistical
moments ${\sigma _{ij}} = \text{Tr}\big[ {{{\{ {{{\hat R}_i},{{\hat R}_j}} \}}_ + }\ {\rho _{AB}}} \big]$.
We can further build a covariance matrix $\sigma_{AB}$ with elements ${\sigma _{ij}}$, which can be written in the standard block form
\begin{equation}\label{w1}
 \sigma_{AB}= \left(
                      \begin{array}{cc}
                        \mathcal{A} & \mathcal{X} \\
                        \mathcal{X}^{T} & \mathcal{B} \\
                      \end{array}
                    \right),
\end{equation}
where $\mathcal{A}$ and $\mathcal{B}$ are the covariance matrices that correspond
to each mode, and $\mathcal{X}$ is the correlation matrix between them. For a physically legitimate Gaussian state, its
covariance matrix must fulfill the uncertainty relation ${\sigma _{AB}} +i\Omega \ge 0$.

Quantum steering denotes a form of nonlocal correlations, which allows one party of a bipartite quantum system to steer (influence) the quantum state of the other party using local measurement. The post-measurement communication between the parties occurs for detecting this effect in compliance with the no-signaling theorem.
According to them,  quantum steering of mode $A$ by mode $B$
of a two-mode Gaussian state can be quantified by the following
measure named  $B\rightarrow A$ quantum steerability \cite{QWE74}
\begin{equation}\label{w3}
{\cal G}^{B \to A}(\sigma_{AB}) =
\mbox{$\max\big\{0, \frac{1}{2} \ln {\frac{\det \mathcal{B}}{\det \sigma_{AB}}}\big\}$}.
\end{equation}
Similarly, we have the quantification of $A\rightarrow B$ Gaussian steerability
\begin{equation}\label{w4}
{\cal G}^{A \to B}(\sigma_{AB}) =
\mbox{$\max\big\{0, \frac{1}{2} \ln {\frac{\det \mathcal{A}}{\det \sigma_{AB}}}\big\}$}.
\end{equation}
The matrix $\mathcal{A}$ and $\mathcal{B}$ in the above two equations are defined in Eq.(\ref{w1}).

Unlike quantum entanglement,  quantum steering has an asymmetric property. We can distinguish it into three cases: (i) no-way steering: ${\cal G}^{A \to B}(\sigma_{AB})=0$ and ${\cal G}^{B \to A}(\sigma_{AB})=0$; (ii) one-way steering: ${\cal G}^{A \to B}(\sigma_{AB})>0$ and ${\cal G}^{B \to A}(\sigma_{AB})=0$ or  ${\cal G}^{A \to B}(\sigma_{AB})=0$ and  ${\cal G}^{B \to A}(\sigma_{AB})>0$; (iii) two-way steering: ${\cal G}^{A \to B}(\sigma_{AB})>0$ and ${\cal G}^{B \to A}(\sigma_{AB})>0$. The second case demonstrates clearly
the asymmetric nature of quantum correlations which is conjectured to play an important role in various communication protocols \cite{L4,L5,L6,L7,L8,L9}.
To check the degree of  asymmetric steerability,  one can define the steering asymmetry as
\begin{equation}\label{w5}
{\cal G}_{AB}^\bigtriangleup=|{\cal G}^{A \to B}(\sigma_{AB})-{\cal G}^{B \to A}(\sigma_{AB})|.
\end{equation}
The steering asymmetry ${\cal G}_{AB}^\bigtriangleup$ for two-mode Gaussian states cannot exceed $\ln 2 $.

\section{ Gravitational effect as the Gaussian  channels \label{GSCDGE}}
The  Schwarzschild-de Sitter (SdS) spacetime has line element given by
\begin{equation}\label{w6}
\begin{aligned}
ds^{2}=-f(r)dt^{2}+f(r)^{-1}dr^{2}+r^{2}\big(d\theta^{2}+\sin^{2}\theta d\phi^{2}\big),
\end{aligned}
\end{equation}
where $f(r)=1-\frac{2M}{r}-\frac{\Lambda r^{2}}{3}$ with the mass $M$  of the black hole and the cosmological constant $\Lambda$ \cite{QWE75}.
For $3M\sqrt{\Lambda}<1$, this SdS spacetime admits three Killing horizons
\begin{equation}\label{w7}
\begin{aligned}
r_{H}=\frac{2}{\sqrt{\Lambda}}\cos\bigg[\frac{\pi+\cos^{-1}(3M\sqrt{\Lambda})}{3}\bigg],
r_{C}=\frac{2}{\sqrt{\Lambda}}\cos\bigg[\frac{\cos^{-1}(3M\sqrt{\Lambda})-\pi}{3}\bigg],
r_{U}=-(r_{H}+r_{C}),
\end{aligned}
\end{equation}
where $r_{H}$ and $r_{C}$ are respectively the BEH and the CEH of the SdS spacetime. Here, the solution $r_{U}$ is negative and unphysical because we cannot extend the coordinate range beyond the curvature singularity at $r=0$.
Generally, we have $r_{H}<r_{C}$. For $3M\sqrt{\Lambda}\rightarrow1$, we obtain $r_{H}\rightarrow r_{C}$, which is known as the Nariai limit.
The surface gravities of the black hole and an expanding universe are, respectively, expressed as
\begin{equation}\label{w8}
\begin{aligned}
\kappa_{H}=\frac{\Lambda (2r_{H}+r_{C})(r_{C}-r_{H})}{6r_{H}},\quad -\kappa_{C}=\frac{\Lambda (2r_{C}+r_{H})(r_{H}-r_{C})}{6r_{C}}.
\end{aligned}
\end{equation}
Since the repulsive effects are generated by a positive cosmological constant, the surface gravity of an expanding  universe is negative. The two event horizons of the SdS spacetime generate two thermodynamic relationships with temperatures $\frac{\kappa_{H}}{2\pi}$ and $\frac{\kappa_{C}}{2\pi}$. Because $r_{H}<r_{C}$, we have $\kappa_{H}>\kappa_{C}$.
This means that the Hawking temperature of the black hole $\frac{\kappa_{H}}{2\pi}$
 is always bigger than the Hawking temperature of an expanding universe $\frac{\kappa_{C}}{2\pi}$.

Because $r=r_{H}, r_{C}$ are two coordinate singularities \cite{QWE70}, we consider Kruskal-like coordinates in order to extend the spacetime beyond them
\begin{equation}\label{w9}
\begin{aligned}
ds^{2}=-\frac{2M}{r}\big|1-\frac{r}{r_{C}}\big|^{1+\frac{\kappa_{H}}{\kappa_{C}}}\big(1+\frac{r}{r_{H}+r_{C}}\big)^{1-\frac{\kappa_{H}}{\kappa_{U}}}
d\bar{\mu}_{H}d\bar{\nu}_{H}+r^{2}(d\theta^{2}+\sin^{2}\theta d\phi^{2}),
\end{aligned}
\end{equation}
\begin{equation}\label{w10}
\begin{aligned}
ds^{2}=-\frac{2M}{r}\big|\frac{r}{r_{H}}-1\big|^{1+\frac{\kappa_{C}}{\kappa_{H}}}\big(1+\frac{r}{r_{H}+r_{C}}\big)^{1+\frac{\kappa_{C}}{\kappa_{U}}}
d\bar{\mu}_{C}d\bar{\nu}_{C}+r^{2}(d\theta^{2}+\sin^{2}\theta d\phi^{2}),
\end{aligned}
\end{equation}
where
\begin{equation}\label{w11}
\begin{aligned}
\bar{\mu}_{H}=-\frac{1}{\kappa_{H}}e^{-\kappa_{H}\mu}, \quad \bar{\nu}_{H}=\frac{1}{\kappa_{H}}e^{\kappa_{H}\nu}, \quad
\bar{\mu}_{C}=\frac{1}{\kappa_{C}}e^{\kappa_{C}\mu}, \quad
\bar{\nu}_{C}=-\frac{1}{\kappa_{C}}e^{-\kappa_{C}\nu},
\end{aligned}
\end{equation}
are the Kruskal null coordinates. Note that $\mu=t-r_{\star}$ and $\nu=t+r_{\star}$ are the usual retarded and advanced null coordinates with the radial tortoise coordinate, which is written as
\begin{equation}\label{w12}
\begin{aligned}
r_{\star}=\frac{1}{2\kappa_{H}}\ln\big|\frac{r}{r_{H}}-1\big|-\frac{1}{2\kappa_{C}}\ln\big|1-\frac{r}{r_{C}}\big|
+\frac{1}{2\kappa_{U}}\ln\big|\frac{r}{r_{U}}-1\big|.
\end{aligned}
\end{equation}
One finds  that Eqs.(\ref{w9}) and (\ref{w10}) are free of coordinate singularities, respectively, on BEH and CEH of the SdS spacetime. However, there is no single Kruskal coordinate for the SdS spacetime, which simultaneously removes the coordinate singularities for both horizons.
The Kruskal timelike and spacelike coordinates can be defined  as
\begin{equation}\label{w13}
\begin{aligned}
\bar{\mu}_{H}=T_{H}-R_{H},\quad
\bar{\nu}_{H}=T_{H}+R_{H},\quad
\bar{\mu}_{C}=T_{C}-R_{C},\quad
\bar{\nu}_{C}=T_{C}+R_{C}.
\end{aligned}
\end{equation}
By using Eq.(\ref{w11}), we, respectively, obtain the relationships
\begin{equation}\label{w14}
\begin{aligned}
-\bar{\mu}_{H}\bar{\nu}_{H}&=R^{2}_{H}-T^{2}_{H}=\frac{1}{\kappa^{2}_{H}}\big|1-\frac{r}{r_{C}}\big|^{-\kappa_{H}/\kappa_{C}}
\big|\frac{r}{r_{U}}-1\big|^{\kappa_{H}/\kappa_{C}}(\frac{r}{r_{H}}-1),\\
-\bar{\mu}_{C}\bar{\nu}_{C}&=R^{2}_{C}-T^{2}_{C}=-\frac{1}{\kappa^{2}_{C}}\big|\frac{r}{r_{U}}-1\big|^{-\kappa_{C}/\kappa_{U}}
\big|\frac{r}{r_{H}}-1\big|^{-\kappa_{C}/\kappa_{H}}(1-\frac{r}{r_{C}}).
\end{aligned}
\end{equation}

\begin{figure}
\includegraphics[scale=0.9]{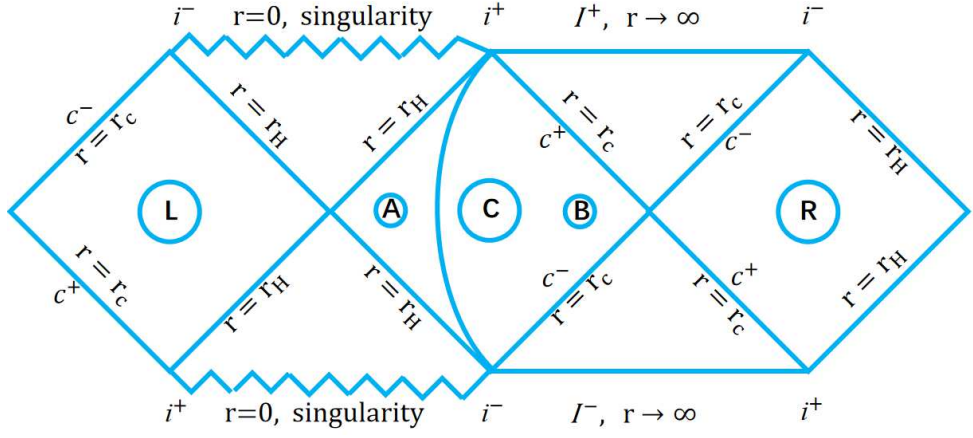}
\caption{ The Penrose-Carter diagram plots the causal structure of the extended SdS spacetime. $i^{\pm}$ respectively denote the future and past timelike infinities and the infinities $ I^{\pm}$ are spacelike. A thermally opaque membrane placed in region $C$ ($r_H< r < r_C$)  cuts it into two subregions: $A$ and $B$. The regions $R$, $L$ are time reversed with respect to the region $C$. All the seven wedges are causally disconnected. Alice and Bob are restricted to the subregions $A$ and $B$, respectively. }\label{Fig1}
\end{figure}

According to Eq.(\ref{w8}),  any equilibrium of two types of temperature
is not possible.  One way to solve this problem is to place a
thermally opaque membrane in the region $C$, which separates the region $C$ into two thermally isolated subregions $A$ and $B$ ($C=A\cup B$) in Fig.\ref{Fig1} \cite{QWE70,QWE75}. Therefore, an observer located at the BEH detects Hawking radiation at temperature $\frac{\kappa_{H}}{2\pi}$, and another observer located at the CEH detects Hawking radiation at another temperature $\frac{\kappa_{C}}{2\pi}$.
Note analogically that even in an asymptotically flat black hole spacetime, a perfectly thermally reflecting membrane is needed to encase the black hole in order to define the Hattel-Hawking state, which describes the thermal equilibrium of the black hole with blackbody radiation at the Hawking temperature \cite{QWE77}.

We consider a free, massless, and minimally coupled scalar field that satisfies the Klein-Gordon equation \cite{QWE78}
\begin{eqnarray}\label{w15}
\frac{1}{\sqrt{-g}}\frac{\partial}{\partial x^{\mu}}(\sqrt{-g}g^{\mu\nu}\frac{\partial\Psi}{\partial x^{\nu}})=0.
\end{eqnarray}
For any of the coordinate systems, the mode functions by solving Eq.(\ref{w15}) are simply plane waves.  Firstly, we consider the side of the membrane that
faces the black hole in subregion $A$. Similar to the Unruh effect, the field quantization can be done in a similar method. Therefore, we
can use the black hole mode and the Kruskal mode to quantize the scalar field, respectively,
and then obtain the Bogoliubov transformations between the creation and annihilation operators
in different coordinates \cite{QWE78}. After properly normalizing the state vector, the Kruskal
vacuum becomes
\begin{equation}\label{w16}
\begin{aligned}
|0\rangle_{\kappa_{H}}=\sum_{n=0}^{\infty}\frac{\tanh^{n}r}{\cosh r}|n_{A},n_{L}\rangle,
\end{aligned}
\end{equation}
where $\cosh r=(1-e^{-\frac{2\pi\omega}{\kappa_{H}} })^{-\frac{1}{2}}$, $|n\rangle_{A}$ and $|n\rangle_{L}$ are the orthogonal bases in the regions $A$ and $L$, respectively.
Here, the region $L$ is causally disconnected the region $A$.
Similarly,  the Kruskal vacuum in an expanding universe can be given by
\begin{equation}\label{w17}
\begin{aligned}
|0\rangle_{\kappa_{C}}=\sum_{n=0}^{\infty}\frac{\tanh^{n}w}{\cosh w}|n_{B},n_{R}\rangle,
\end{aligned}
\end{equation}
where $\cosh w=(1-e^{-\frac{2\pi\omega}{\kappa_{C}} })^{-\frac{1}{2}}$, $|n\rangle_{B}$ and $|n\rangle_{R}$ are the orthogonal bases in the regions $B$ and $R$, respectively.
Here, the regions $R$ and $B$  are causally disconnected.
Note that although the construction of the Gibbons-Hawking state is mathematically consistent, such a  thermally opaque membrane can actually be achieved between BEH and CEH. Perhaps one possible way to achieve this is to consider the Klein-Gordon equation with the radial function that satisfies
\begin{eqnarray}\label{w3909}
\displaystyle{\left(-\frac{\partial^{2}}{\partial t^{2}}+\frac{\partial^{2}}{\partial r_{\star}^{2}}\right)}{R}(r)+\displaystyle{\left(1-\frac{2M}{r}-\frac{\Lambda r^{2}}{3}\right)}\displaystyle{\left(\frac{l(l+1)}{r^{2}}+\frac{2M}{r^{3}}-\frac{\Lambda}{3}\right)}R(r)=0.
\end{eqnarray}
The effective potential term that appears in the above equation disappears at both BEH and CEH and is positive in between them. Therefore, This bell shaped potential will serve as a barrier between BEH and CEH. Modes that cannot penetrate it will be limited to the regions near the horizons and will be separated from each other. Hence, the effective potential can be considered a natural realization of the thermally opaque membrane described above.

From  Eqs.(\ref{w16}) and (\ref{w17}), we can see that the effect of gravitational effect can be described by the two-mode squeezing operators $U_{AL}$ and $U_{BR}$ \cite{Q40}. The two-mode squeezing operators in phase space are the Gaussian operations that preserve the Gaussianity of the input states, and correspond to the symplectic transformations
\begin{eqnarray}\label{w18}
 S_{A,L}(r)= \left(\!\!\begin{array}{cc}
\cosh r I_{2}&\sinh r Z_{2}\\
\sinh r Z_{2}&\cosh r I_{2}
\end{array}\!\!\right),
\end{eqnarray}
\begin{eqnarray}\label{w19}
S_{B,R}(w)= \left(\!\!\begin{array}{cc}
\cosh w I_{2}&\sinh w Z_{2}\\
\sinh w Z_{2}&\cosh w I_{2}
\end{array}\!\!\right).
\end{eqnarray}
Here, $I_{2}$ is the $2\times2$ identity matrix and $Z_2$ is the Pauli $Z$ operator.

\section{ Gaussian quantum steering in multi-event horizon SdS spacetime \label{GSCDGE}}
In this paper, we initially consider a two-mode squeezed Gaussian Kruskal state with squeezing
parameter $s$ shared by Alice and Bob \cite{Q40}. Its covariance matrix can be expressed  as
\begin{eqnarray}\label{w20}
\sigma_{AB}(s)= \left(\!\!\begin{array}{cc}
\cosh 2s I_{2}&\sinh 2s Z_{2}\\
\sinh 2s Z_{2}&\cosh 2s I_{2}
\end{array}\!\!\right).
\end{eqnarray}
Here, the mode $A$ observed by Alice is located at the BEH in subregion $A$ of $C$, and  the mode $B$ observed by Bob is located at
the CEH in subregion $B$ of $C$.
Due to this placement, the gravitational effect corresponds to the two-mode squeezing operations related to the symplectic transformations $ S_{A,L}(r)$ and $S_{B,R}(w)$. Under such transformations, the initial modes  $A$ and $B$ are  mapped into four sets of modes: the mode $A$ in region $A$; the mode $B$ in region $B$; the mode $\bar A$ in region $L$; the mode $\bar B$ in region $R$. Therefore, the covariance matrix that describes the complete system is given by
\begin{eqnarray}\label{w21}
\nonumber\sigma_{AB \bar A \bar B}(s,r,w) &=& \big[S_{A,\bar A}(r) \oplus  S_{B,\bar B}(w)\big] \big[\sigma_{AB}(s) \oplus I_{\bar A\bar B}\big]\\&& \big[S_{A,\bar A}(r) \oplus  S_{B,\bar B}(w)\big]\,^{\sf T},
\end{eqnarray}
where $S_{A,\bar A}(r)$ and $S_{B,\bar B}(w)$ for Alice and Bob are the symplectic
phase space representations of two-mode squeezing operation, which are given by Eqs.(\ref{w18}) and (\ref{w19}).

Because the regions $A$ and $B$ are causally disconnected from the regions $L$ and $R$, respectively, we should take the trace over modes $\bar A$ and $\bar B$. By performing this operation on Eq.(\ref{w21}), we obtain the covariance matrix $\sigma_{AB }(s,r,w)$ between Alice and Bob
\begin{equation}\label{w22}
\sigma_{AB}(s,r,w)= \left( {\begin{array}{*{20}{c}}
\mathcal{A}_{AB} & \mathcal{X}_{AB}\\
{{\mathcal{X}_{AB}^{\sf T}}} & \mathcal{B}_{AB}\\
\end{array}} \right),
\end{equation}
where $\mathcal{A}_{AB}=[\cosh(2s) \cosh^2(r) + \sinh^2(r)]I_2$, $\mathcal{X}_{AB}=[\cosh(r) \cosh(w)\sinh(2s)]Z_2$ and $\mathcal{B}_{AB}=[\cosh(2s) \cosh^2(w) + \sinh^2(w)]I_2$.
Employing Eqs.(\ref{w3}) and (\ref{w4}), we obtain the analytic expressions for the Gaussian quantum steering $B\rightarrow A$ and $A\rightarrow B$
 \begin{equation}\label{w23}
{\cal G}^{B \to A}(\sigma_{AB}) =
\mbox{$\max\big\{0,  \ln {\frac{\cosh(2s) \cosh^2 w + \sinh^2 w}{\mathbf{Z}}}\big\}$},
\end{equation}
\begin{equation}\label{w25}
{\cal G}^{A \to B}(\sigma_{AB}) =
\mbox{$\max\big\{0,  \ln {\frac{\cosh(2s) \cosh^2 r + \sinh^2 r}{\mathbf{Z}}}\big\}$},
\end{equation}
where  $\mathbf{Z}=\cosh(2s)[\cosh^{2} r\sinh^{2} w + \sinh^{2} r\cosh^{2} w]+\cosh^{2} r\cosh^{2} w+\sinh^{2} r\sinh^{2}w$.
We can see that the Gaussian quantum steering depends not only on the squeezing parameter $s$, but also on the effects of the black hole and an expanding universe.
The Gaussian steering asymmetry is thus calculated via Eq.(\ref{w5}).

\begin{figure}
\begin{minipage}[t]{0.5\linewidth}
\centering
\includegraphics[width=3.0in,height=5.2cm]{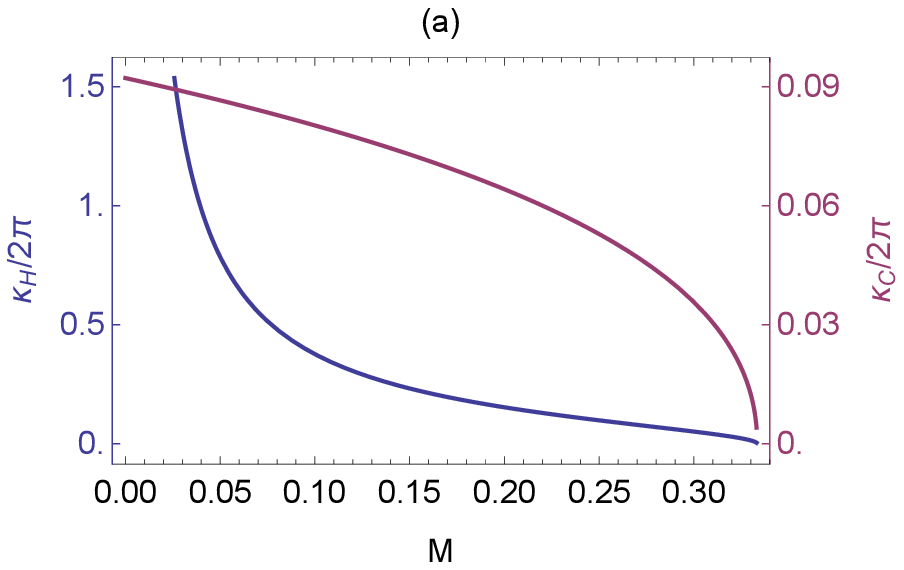}
\label{fig1a}
\end{minipage}%
\begin{minipage}[t]{0.5\linewidth}
\centering
\includegraphics[width=3.0in,height=5.2cm]{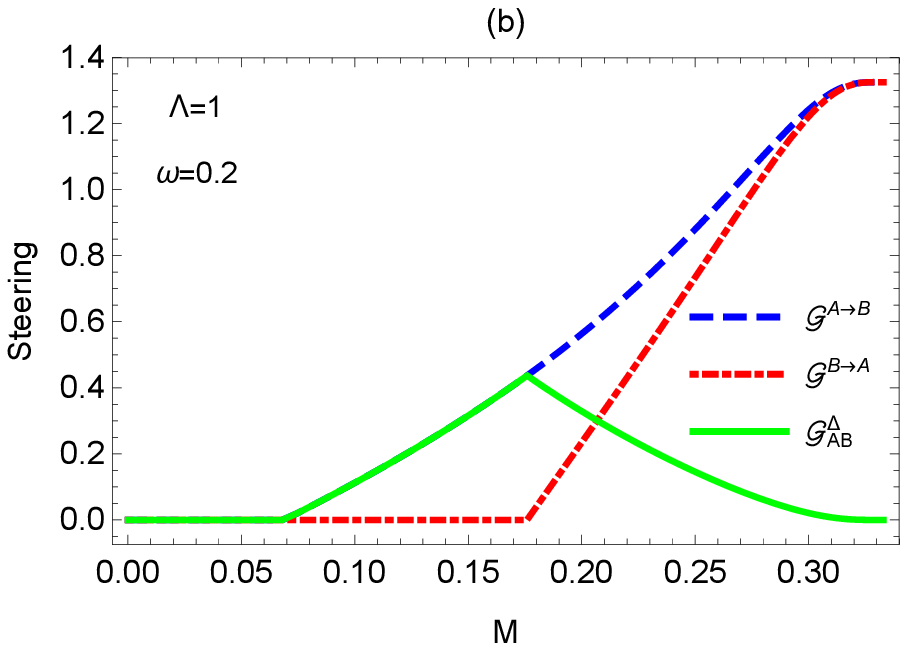}
\label{fig1c}
\end{minipage}%
\caption{The Hawking temperature $\frac{\kappa_{H}}{2\pi}$ of BEH, the Hawking temperature $\frac{\kappa_{C}}{2\pi}$ of CEH, the Gaussian quantum steering ${\cal G}^{A \to B}$, ${\cal G}^{B \to A}$ and their steering asymmetry ${\cal G}^\Delta_{AB}$ as functions of the mass $M$ of the black hole for fixed $\omega=0.2$ and $\Lambda=s=1$.}
\label{Fig2}
\end{figure}

The dependency of the Hawking temperatures $\frac{\kappa_{H}}{2\pi}$ of BEH and $\frac{\kappa_{C}}{2\pi}$ of CEH on the mass $M$ of the black hole is given in Fig.\ref{Fig2}(a). It is shown that both $\frac{\kappa_{H}}{2\pi}$ and $\frac{\kappa_{C}}{2\pi}$ are the monotonically decreasing functions of the mass $M$ of the black hole.
When the mass $M$ of the black hole decreases, the gravitational effect becomes stronger and stronger, thus quantum steering decreases continuously, and eventually both ${\cal G}^{A \to B}$ and ${\cal G}^{B \to A}$ happen sudden death [see Fig.\ref{Fig2}(b)]. Due to the different $M$ values of sudden death, quantum steering transforms from two-way to one-way and no-way respectively. Interestingly, quantum steering ${\cal G}^{A \to B}$  is always bigger than quantum steering ${\cal G}^{B \to A}$ before sudden death. Note that the surface gravity $\kappa_H$ is always bigger than the surface gravity $\kappa_C$, thus the result is nontrivial and nonintuitive, which means that the observer who has larger temperature has stronger steerability than the other one. This result also can be verified though Eqs.(\ref{w23}) and (\ref{w25}) by observing that $\cosh(2s) \cosh^2(r) + \sinh^2(r)>\cosh(2s) \cosh^2(w) + \sinh^2(w)$.

If a two-mode Gaussian state is immersed in the two independent thermal baths at an asymptotically flat region, then a similar conclusion could also be correct \cite{QQWE78}. For example,  for two fixed temperatures of the thermal bath ($T_1$ and $T_2$), Charlie who stays in the thermal bath with a higher temperature $T_1$  ($T_1> T_2$) has more stronger steerability from Charlie to David than quantum steerability from David to Charlie. Therefore, the phenomenon  is not a genuine effect of SdS spacetime. The clarification needed is that as long as one or both of the fixed temperatures of the thermal bath increase ($T_3> T_1$ and $T_4\geq T_2$ ), quantum steerability from Charlie to David and quantum steerability from David to Charlie decrease.  In addition, the difference is that the Hawking effect of the expanding universe in  multi-event horizon spacetime may increase quantum steering (see below), while the thermal bath can only reduce quantum steering.

We also show the steering asymmetry ${\cal G}^\Delta_{AB}$ as a function of $M$, which reveals a non-monotonic change in $M$. The  steering asymmetry for  two-way steering is determined by
$${\cal{G}}^{\Delta,2}_{AB}=\ln\frac{\cosh (2s)\cosh^{2}r+\sinh^{2}r}{\cosh(2s)\cosh^{2}w+\sinh^{2}w},$$ which is a decreasing function of the $M$. The  steering asymmetry for  one-way steering is given by
$${\cal{G}}^{\Delta,1}_{AB}=\ln\frac{\cosh (2s)\cosh^{2}r+\sinh^{2}r}{\mathbf{Z}},$$ which is an increasing function of the $M$. The maximal steering asymmetry takes place at the transition point from two-way steering to one-way steering.

\begin{figure}
\begin{minipage}[t]{0.5\linewidth}
\centering
\includegraphics[width=3.0in,height=5.2cm]{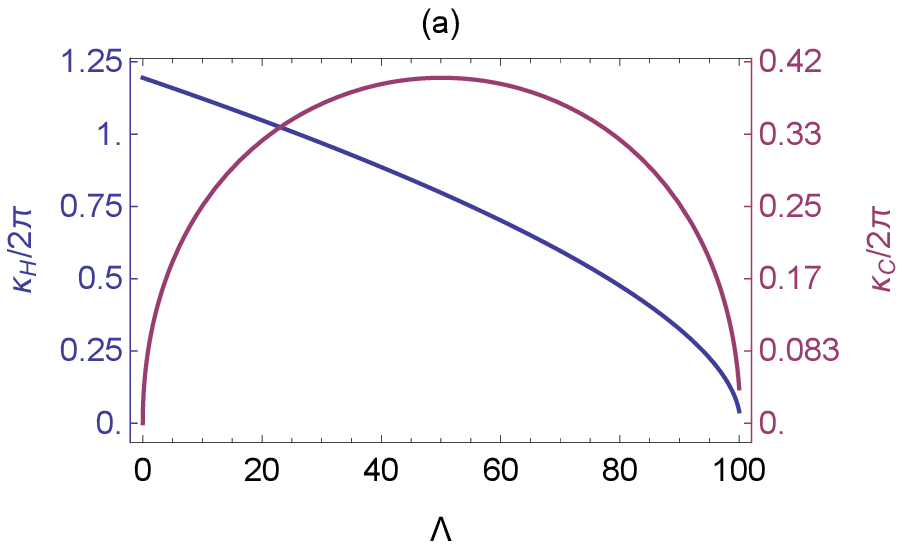}
\label{fig1a}
\end{minipage}%
\begin{minipage}[t]{0.5\linewidth}
\centering
\includegraphics[width=3.0in,height=5.2cm]{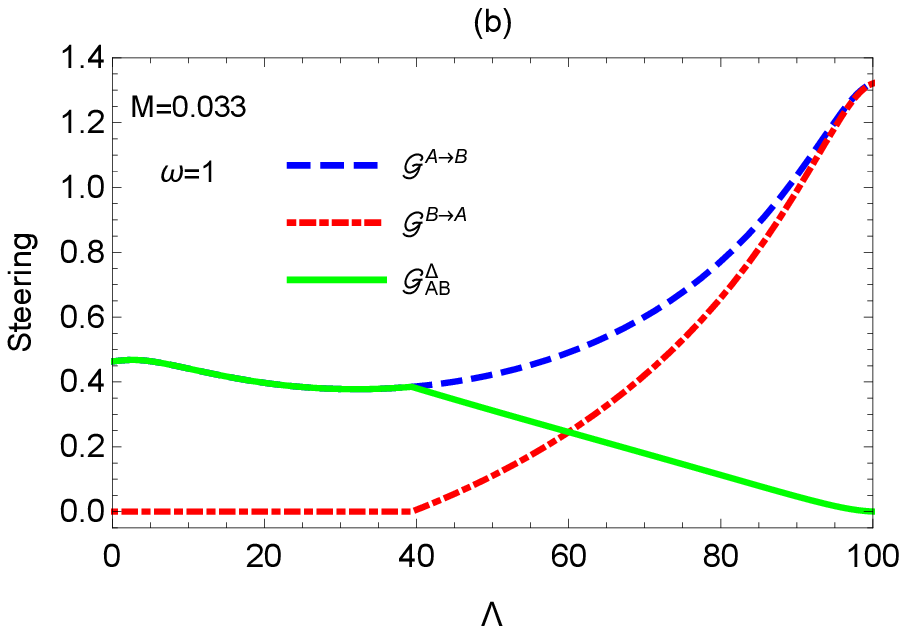}
\label{fig1c}
\end{minipage}%
\caption{The Hawking temperature $\frac{\kappa_{H}}{2\pi}$ of BEH, the Hawking temperature $\frac{\kappa_{C}}{2\pi}$ of CEH, the Gaussian quantum steering ${\cal G}^{A \to B}$, ${\cal G}^{B \to A}$ and steering asymmetry ${\cal G}^\Delta_{AB}$ as functions of the cosmological constant $\Lambda$ for fixed $\omega=s=1$ and $M=0.033$.}
\label{Fig3}
\end{figure}

In Fig.\ref{Fig3}(a), we show the dependency of the Hawking temperatures  $\frac{\kappa_{H}}{2\pi}$ and $\frac{\kappa_{C}}{2\pi}$ on the cosmological constant $\Lambda$. It is shown that with the increase of the cosmological constant $\Lambda$,
the Hawking temperature $\frac{\kappa_{H}}{2\pi}$ of BEH decreases monotonically, while the Hawking temperature $\frac{\kappa_{C}}{2\pi}$ of CEH changes non-monotonically.
In Fig.\ref{Fig3}(b), we plot the Gaussian quantum steering ${\cal G}^{A \to B}$, ${\cal G}^{B \to A}$, and steering asymmetry ${\cal G}^\Delta_{AB}$
as functions of the cosmological constant $\Lambda$.
It is shown that, with the increase of the cosmological constant $\Lambda$, the steering ${\cal G}^{B \to A}$ experiences a sudden birth at some $\Lambda$ and then increases monotonically, eventually approaches to the initial value $\ln[\cosh(2s)]$ in the Nariai limit $3M\sqrt{\Lambda}\rightarrow 1$. For $40<\Lambda<50$, the Hawking temperature $\frac{\kappa_{C}}{2\pi}$ of CEH increases with the growth of  the $\Lambda$, and the steering ${\cal G}^{A \to B}$ and the steering ${\cal G}^{B \to A}$  increase with the $\Lambda$. This means that the Hawking effect the expanding universe can enhance and protect quantum steering in the SdS  spacetime. Combining Fig.\ref{Fig2} and Fig.\ref{Fig3}, we can conclude that the Hawking temperature $\frac{\kappa_{H}}{2\pi}$ of BEH downgrades Gaussian quantum steering monotonously, while the Hawking temperature $\frac{\kappa_{C}}{2\pi}$ of CEH changes Gaussian quantum steering non-monotonically.

 \begin{figure}[htbp]
\centering
\includegraphics[height=1.8in,width=2.0in]{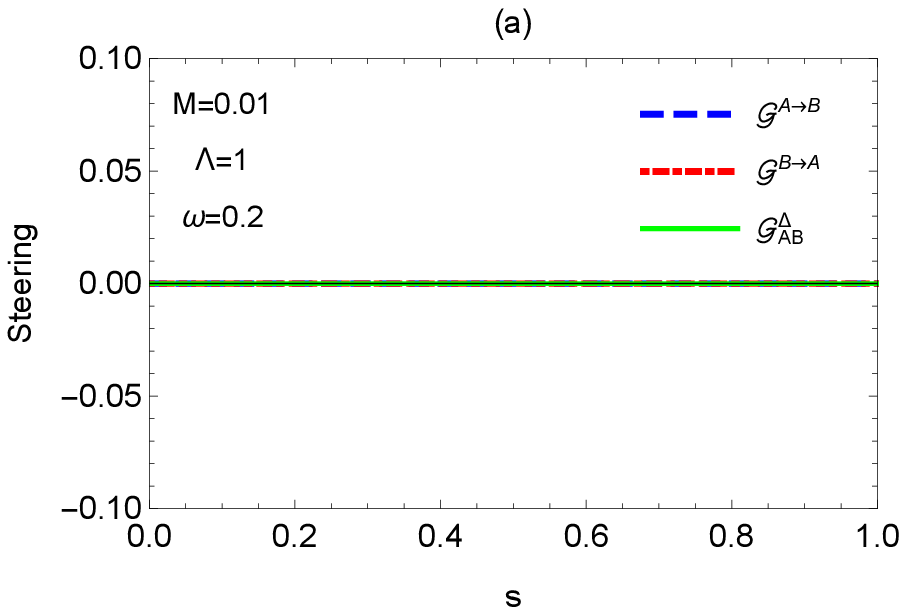}
\includegraphics[height=1.8in,width=2.0in]{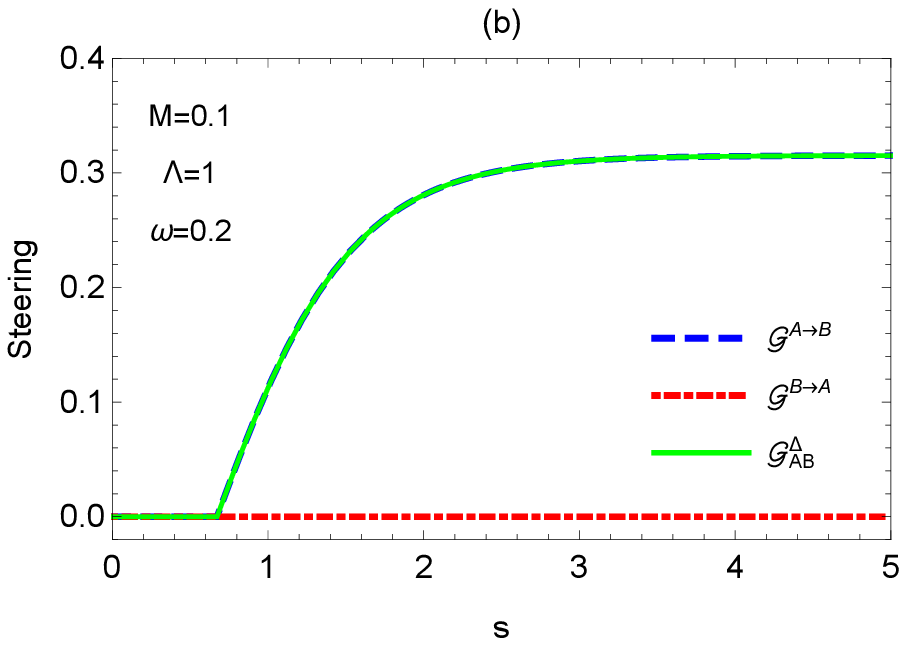}
\includegraphics[height=1.8in,width=2.0in]{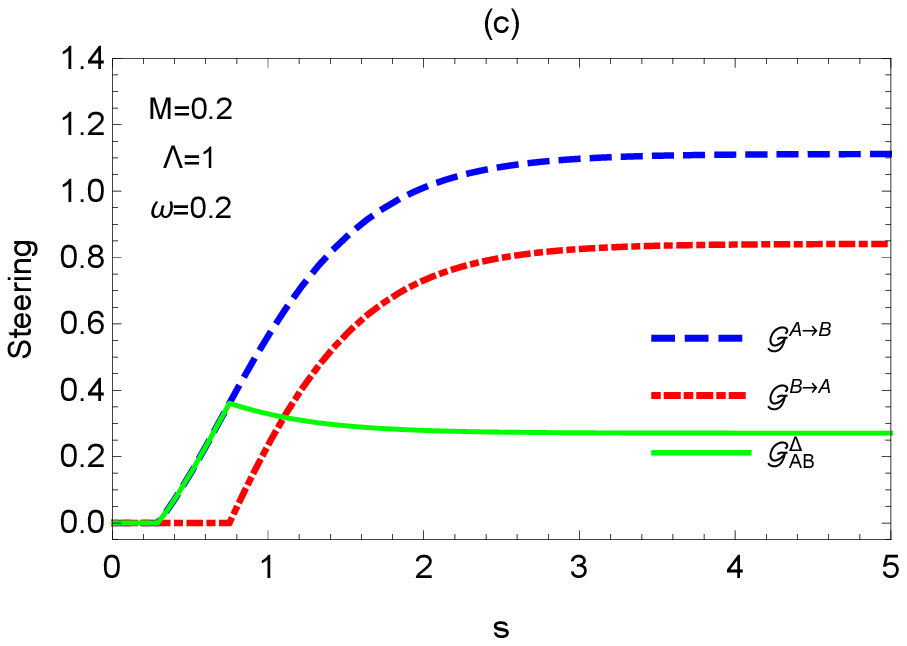}
\caption{ Gaussian quantum steering ${\cal G}^{A \to B}$, ${\cal G}^{B \to A}$ and steering asymmetry ${\cal G}^\Delta_{AB}$ as functions of the squeezing parameter  $s$ for different  mass $M$ of the black hole and fixed $\Lambda=1$ and $\omega=0.2$.}\label{Fig4}
\end{figure}

In Fig.\ref{Fig4},  we plot Gaussian quantum steering and the steering asymmetry as functions of the squeezing parameter  $s$ for different mass $M$ of the black hole. We see that, for $M=0.01$ [Fig.\ref{Fig4}(a)], quantum steering cannot be generated no matter how large the squeezing parameter $s$ is; For $M=0.1$ [Fig.\ref{Fig4}(b)], quantum steering ${\cal G}^{A \to B}$ appears sudden birth at some $s$
and then increases monotonically with $s$, while steering ${\cal G}^{B \to A}$ remains zero. The steering asymmetry ${\cal{G}}^{\Delta}_{AB}$ is determined by ${\cal G}^{A \to B}$ in this case; For $M=0.2$ [Fig.\ref{Fig4}(c)], both ${\cal G}^{A \to B}$ and ${\cal G}^{B \to A}$ appear sudden birth and then increase monotonically with $s$. In this case, ${\cal G}^{A \to B}$ is always larger than ${\cal G}^{B \to A}$, and the maximal steering asymmetry ${\cal{G}}^{\Delta}_{AB}$ indicates the transition from one-way to two-way steering. These results demonstrate the fact that entanglement is the necessary but not sufficient condition of quantum steering and quantum steering becomes feasible only when entanglement reaches some levels.

\begin{figure}
\begin{minipage}[t]{0.5\linewidth}
\centering
\includegraphics[width=3.0in,height=5.2cm]{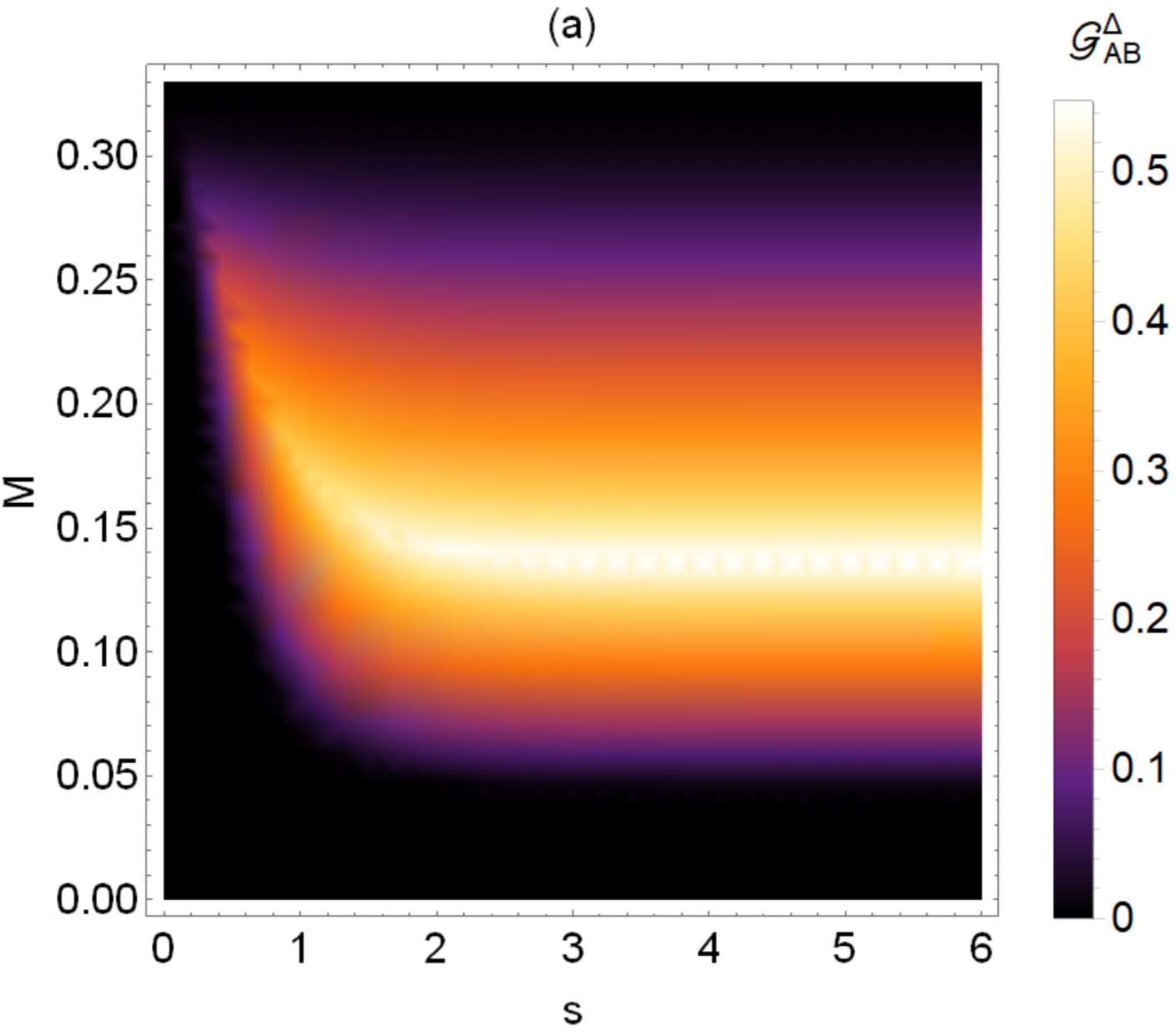}
\label{fig1a}
\end{minipage}%
\begin{minipage}[t]{0.5\linewidth}
\centering
\includegraphics[width=3.0in,height=5.2cm]{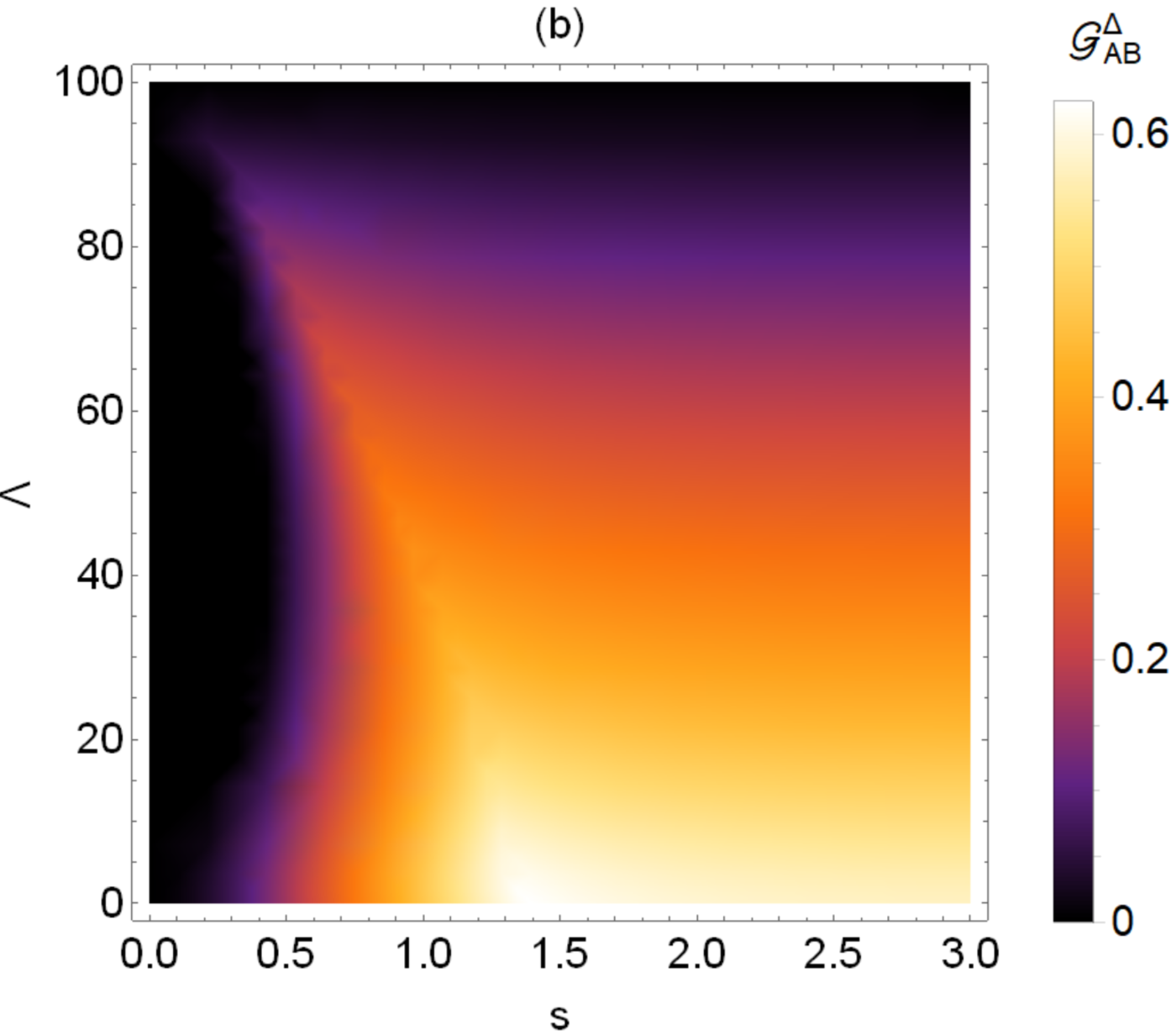}
\label{fig1c}
\end{minipage}%
\caption{(a) Gaussian steering asymmetry ${\cal G}^\Delta_{AB}$ as functions of the squeezing parameter $s$ and the mass $M$ of the black hole for fixed  $\Lambda=1$ and $\omega=0.2$. (b) Gaussian steering asymmetry ${\cal G}^\Delta_{AB}$ as functions of the squeezing parameter $s$ and the cosmological constant $\Lambda$ for fixed $M=0.033$ and $\omega=1$.}
\label{Fig5}
\end{figure}

To better understand the interplay between squeezing and the gravitational effect of SdS spacetime in the generation of  Gaussian
steering, we plot Gaussian steering asymmetry ${\cal G}^\Delta_{AB}$
as a function of $s$ and $M$ or of $s$ and $\Lambda$ in Fig.\ref{Fig5}.  We see from Fig.\ref{Fig5}(a) that, for smaller ($M<0.05$) and larger ($M>0.3$) mass of the black hole, steering asymmetry ${\cal G}^\Delta_{AB}$ remains zero no matter how large the squeezing parameter $s$ is, in which both ${\cal G}^{A \to B}$ and ${\cal G}^{B \to A}$ are zero indeed [see Fig.\ref{Fig4}(a)]; For the middle $M$ (say $M=0.15$), steering asymmetry ${\cal G}^\Delta_{AB}$ increases from zero and then remains almost an asymptotic value  with the increasing of $s$, which corresponds indeed the case of Fig.\ref{Fig4}(c); The purple-yellow regions (say $M=0.1$ and 0.25) correspond to the case of Fig.\ref{Fig4}(b), in which only one steering (i.e. ${\cal G}^{A \to B}$) occurs sudden birth. Fig.\ref{Fig5}(b) shows that the maximal steering asymmetry takes place in the region where $\Lambda$ has small values. From these analysis, we conclude that proper values of $M$ and small $\Lambda$ of SdS are beneficial to the asymmetry of Gaussian quantum steering.

\begin{figure}
\includegraphics[scale=0.9]{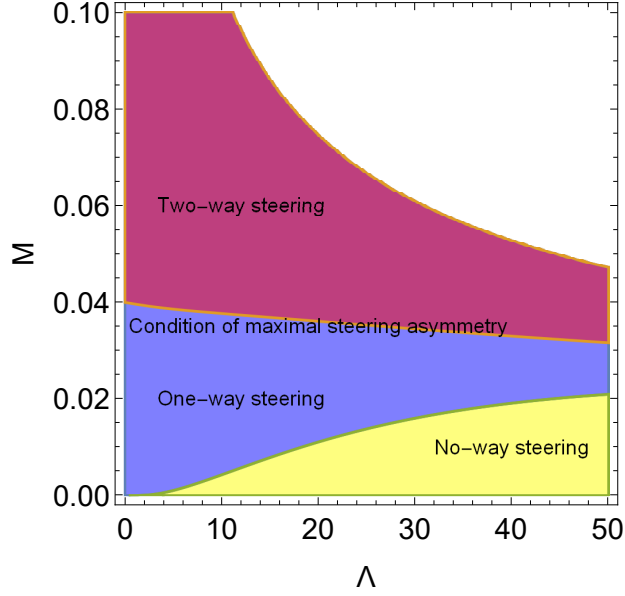}
\caption{Gaussian quantum steering as functions of the cosmological constant $\Lambda$ and  the mass $M$ of the black hole for fixed $\omega=s=1$. }\label{Fig6}
\end{figure}

It is interesting to find the parameter regions  in which two-way,
one-way and no-way quantum steering take place.
Because the Hawking temperature of BEH is always bigger than the Hawking temperature of CEH, resulting in ${\cal G}^{A \to B}\geq{\cal G}^{B \to A}$, the condition for two-way quantum steering (Please refer to the red region in Fig.\ref{Fig6}) is simply determined by ${\cal G}^{B \to A}>0$, or equivalently written as
\begin{eqnarray}\label{condition1}
    \nonumber \cosh (2s)[\cosh^{2}r\sinh^{2}w+\cosh^{2}w(\sinh^{2}r-1)]\\
    +\sinh^{2}w (\sinh^{2}r-1)+\cosh^{2}r\cosh^{2}w<0.
\end{eqnarray}
The condition for one-way quantum steering (Please refer to the blue region in Fig.\ref{Fig6}), ${\cal G}^{A \to B}>0$ and ${\cal G}^{B \to A}=0$, can be written as
\begin{eqnarray}
\left\{
\begin{array}{l}
\cosh (2s)[\cosh^{2}w\sinh^{2}r+\cosh^{2}r(\sinh^{2}w-1)]\\
    +\sinh^{2}r (\sinh^{2}w-1)+\cosh^{2}w\cosh^{2}r<0
\,,\\
\\
\cosh (2s)[\cosh^{2}r\sinh^{2}w+\cosh^{2}w(\sinh^{2}r-1)]\\
    +\sinh^{2}w (\sinh^{2}r-1)+\cosh^{2}r\cosh^{2}w\geq0
\,.
\label{solutions}
\end{array}
\right.
\end{eqnarray}
Finally, the condition for no-way quantum steering (Please refer to the yellow region in Fig.\ref{Fig6}), ${\cal G}^{A \to B}=0$, is given by
\begin{eqnarray}\label{condition11}
    \nonumber \cosh (2s)[\cosh^{2}r\sinh^{2}w+\cosh^{2}w(\sinh^{2}r-1)]\\
    +\sinh^{2}w (\sinh^{2}r-1)+\cosh^{2}r\cosh^{2}w\geq0.
\end{eqnarray}
The maximal steering asymmetry takes place at the transition from one-way to two-way steering, which fits
\begin{eqnarray}\label{condition12}
    \nonumber \cosh (2s)[\cosh^{2}r\sinh^{2}w+\cosh^{2}w(\sinh^{2}r-1)]\\
    +\sinh^{2}w (\sinh^{2}r-1)+\cosh^{2}r\cosh^{2}w=0.
\end{eqnarray}
We can clearly see the transition between one-way and two-way steering in Fig.\ref{Fig6}. Unfortunately, due to the complexity of calculations, we cannot mathematically provide the scale for the sudden death of quantum steering in SdS spacetime, but we can intuitively draw the region where quantum steering survives in Fig.\ref{Fig6}.

\section{ Gaussian quantum steering under the effective temperature \label{GSCDGE}}
Since the surface gravity of the black hole $\kappa_H$ is bigger than the surface gravity of
an expanding universe $\kappa_C$, the flux of outgoing particles emitted from the BEH
is greater than the flux of particles propagating inward emitted from the CEH at any point $r_H<r<r_C$, which results in an effective outward flux and evaporation of
the black hole \cite{QWE79,QWE80,QWE81}. The effective temperature $T_{\rm{eff}}$ is closely related to these particle fluxes. As is well known, increasing the Hawking temperature of the black hole reduces quantum correlation in the single horizon. However, the influence of effective
temperature on Gaussian quantum steering is unclear in multi-event horizon spacetime.  Therefore, in this section, we will explore how the effective equilibrium temperature $T_{\rm{eff}}$ associated with the Hawking  temperatures $\frac{\kappa_{H}}{2\pi}$ and $\frac{\kappa_{C}}{2\pi}$ affects Gaussian quantum steering.

Since we treat both the horizons together, the analytic Feynman
propagator that connects both the horizons cannot exist \cite{QWE75}. Unlike the single horizon cases, one cannot construct any single global mode that is analytic on both the horizons.
This means that the Kruskal-like coordinates cannot remove the
coordinate singularities of both the horizons.
Therefore, one should introduce a new coordinate system to
solve the issue of the effective temperature $T_{\rm{eff}}$. The BEH  and CEH  for the multi-event spacetime can provide two thermodynamic relationships with the Hawking temperatures $\kappa_H/2\pi$ and $\kappa_C/2\pi$,
\begin{equation}\label{OPw1}
\delta M = \frac{\kappa_H}{2\pi} \frac{\delta A_H}{4}, \qquad \delta M= -\frac{\kappa_C}{2\pi} \frac{\delta A_C}{4},
\end{equation}
where $A_H$ and $A_C$ are the areas of the BEH  and CEH of the SdS spacetime,  respectively. Employing Eqs.(\ref{w7}) and (\ref{w8}), one can define a total entropy $S= (A_{H}+A_C)/4$ from an observer in the entire region $C$, resulting in a thermodynamic relationship to the effective equilibrium temperature $T_{\rm eff}$ in order to treat both
horizons in an equal footing \cite{QWE79,QWE80,QWE81}
\begin{equation}\label{OPw1}
\delta M =- \frac{\kappa_H\kappa_C}{2\pi(\kappa_H-\kappa_C)}\delta S=-T_{\rm eff}\delta S.
\end{equation}
The effective equilibrium  temperature $T_{\rm eff}$ is associated with an emission
probability  corresponding to the particle flux
on the BEH created by CEH. This means that particle creation arises  in a
single region, rather than in causally disconnected spacetime wedges.
Simply, the surface gravity $\kappa_U$ of the unphysical horizon can be expressed as
\begin{equation}\label{w26}
\frac{1}{\kappa_U}=\frac{1}{\kappa_C}-\frac{1}{\kappa_H}.
\end{equation}
We note that the appearance of $\kappa_U$ can guarantee the
existence of the effective equilibrium  temperature $T_{\rm eff}$. From Eqs.(\ref{w8}) and (\ref{w26}), we can find that $T_{\rm eff}\rightarrow 0$ as $\Lambda\rightarrow 0$.

Accordingly, the $t-r$ part of Eq.(\ref{w6}) can be rewritten as
\begin{equation}\label{w27}
\begin{aligned}
ds^{2}=-\frac{2M}{r}\big|1-\frac{r}{r_{C}}\big|^{1+\frac{\kappa_{U}}{\kappa_{C}}}
\big|\frac{r}{r_{H}}-1\big|^{1-\frac{\kappa_{U}}{\kappa_{H}}}d\bar{\mu}d\bar{\nu},
\end{aligned}
\end{equation}
with $\bar{u}=-\frac{1}{\kappa_{U}}e^{-\kappa_{U}\mu}$ and $\bar{\nu}=\frac{1}{\kappa_{U}}e^{\kappa_{U}\nu}$ \cite{QWE70}.
In this way, we analytically extend the spacetime metric at $r_U$. Note that the metric Eq.(\ref{w27}) cannot correspond to the beyond horizon extensions and represents a coordinate system only in the region of interest $A\cup B$, $r_H<r<r_C$.

Now we can use coordinates ($\mu, \nu$) and  ($\bar\mu, \bar\nu$)
to define a field quantization, respectively. After using the standard procedure of field quantization, one obtains the vacuum state $|\bar 0\rangle$ in ($\bar\mu, \bar\nu$) mode \cite{QWE70}
\begin{equation}\label{w29}
\begin{aligned}
|\bar 0\rangle=\sum_{n=0}^{\infty}\frac{\tanh^{n}\gamma}{\cosh \gamma}|n,n\rangle,
\end{aligned}
\end{equation}
which is a squeezed state in SdS spacetime. Here $\cosh \gamma=(1-e^{-\frac{2\pi\omega}{\kappa_{U}} })^{-\frac{1}{2}}$, and the effective temperature is defined by  $T_{\rm{eff}}=\frac{\kappa_U}{2\pi}$.
It should be emphasized that the entangled pair creation is occurring only in region $A\cup B$ ($r_H<r<r_C$).

\begin{figure}
\begin{minipage}[t]{0.5\linewidth}
\centering
\includegraphics[width=3.0in,height=5.2cm]{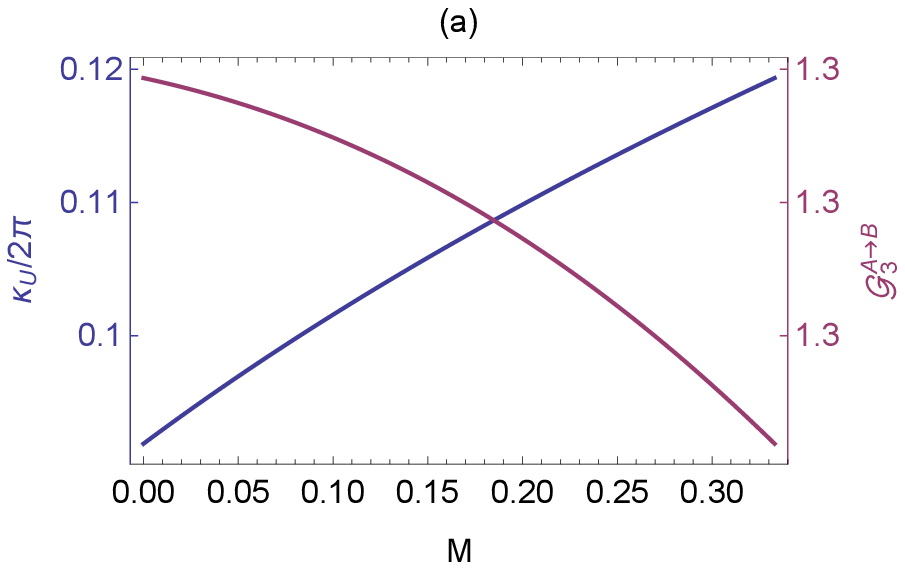}
\label{fig1a}
\end{minipage}%
\begin{minipage}[t]{0.5\linewidth}
\centering
\includegraphics[width=3.0in,height=5.2cm]{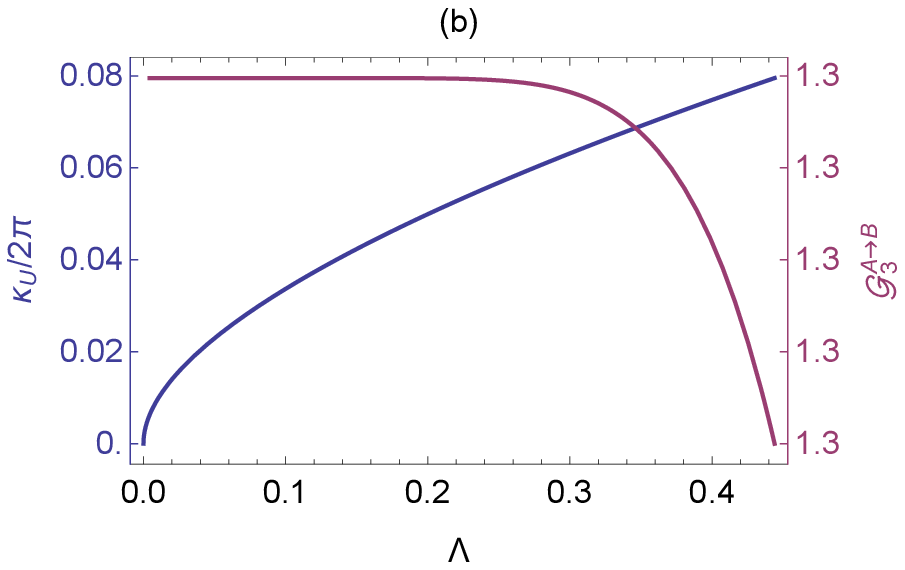}
\label{fig1c}
\end{minipage}%
\caption{The effective equilibrium temperature $\frac{\kappa_{U}}{2\pi}$ and Gaussian quantum steering ${\cal G}^{A \to B}_3$ (a) as a function of the mass $M$ of the black hole for fixed  $\Lambda=s=\omega=1$, and (b) as a function of the cosmological constant $\Lambda$ for fixed $M=0.5$ and $s=\omega=1$.}
\label{Fig7}
\end{figure}

We initially assume that a two-mode squeezed Gaussian state of Eq.(\ref{w20})  is in region $C$, which can be expanded by Eq.(\ref{w29}).
After tedious but straightforward
calculations, we obtain the expressions for Gaussian quantum steering between Alice and Bob
 \begin{equation}\label{w30}
{\cal G}^{A \to B}_3(\sigma_{AB}) ={\cal G}^{B \to A}_3(\sigma_{AB}) =
\mbox{$\max\big\{0,  \ln {\frac{\cosh(2s) \cosh^2\gamma + \sinh^2 \gamma}{\cosh^{2} \gamma[2\cosh(2s)\sinh^{2}\gamma+\cosh^{2}\gamma]+\sinh^{4}\gamma}}\big\}$}.
\end{equation}
It means that the steering asymmetry always vanishes under the scenario of effective temperature.
In Fig.\ref{Fig7},  we plot the effective equilibrium temperature $T_{\rm{eff}}$ and Gaussian quantum steering ${\cal G}^{A \to B}_3$ as functions of the mass $M$ of the black hole
or the cosmological constant $\Lambda$. It shows that the effective equilibrium temperature $T_{\rm{eff}}$ increases  monotonically and accordingly the steering ${\cal G}^{A \to B}_3$ reduces monotonically with the growth of $M$ and $\Lambda$, meaning that ${\cal G}^{A \to B}_3$ is a monotonic decreasing function of $T_{\rm{eff}}$.  It means that from the scenario of effective temperature, the quantum steering never degrades but increases with the increasing of the black hole temperature.
 We emphasize once again that the
(entangled) pair creation with the effective temperature $T_{\rm{eff}}$ only occurs in the single region $r_H<r<r_C$ and not in the causally disconnected wedges
as of the preceding section. Therefore, $|\bar 0\rangle$ cannot be considered any analogue of the global or Minkowski vacuum. We also emphasize that the appearance of the surface gravity $\kappa_U$
(instead of  $\kappa_H$ or  $\kappa_C$) ensures the
emergence of the effective temperature $T_{\rm{eff}}$. Based on the above analysis, the effective temperature $T_{\rm{eff}}$ has completely different meanings with the Hawking  temperatures $\frac{\kappa_{H}}{2\pi}$ and $\frac{\kappa_{C}}{2\pi}$. From Eqs.(\ref{w8}) and (\ref{w26}), we obtain
$$\lim_{3M\sqrt{\Lambda}\rightarrow 0}T_{\rm{eff}}\approx\frac{1}{2\pi}\sqrt{\frac{\Lambda}{3}},\lim_{3M\sqrt{\Lambda}\rightarrow 1}T_{\rm{eff}}\approx\frac{1}{2\pi}\frac{3\sqrt{\Lambda}}{4}.$$ It is to say that the effective temperature $T_{\rm{eff}}$ is not diverging for the extremely hot black hole ($3M\sqrt{\Lambda}\rightarrow 0 $ with $\Lambda$ fixed) and nonvanishing in the Nariai limit ($3M\sqrt{\Lambda}\rightarrow 1$).
With the increase of the Hawking temperature of the black hole due to a decrease in $3M\sqrt{\Lambda}$, the black hole emits more, leading to a greater outward flux on the CEH. This corresponds to a decrease in effective inward flux or a reduction in effective temperature, showing that
increasing the Hawking temperature of the black hole reduces the effective temperature. Therefore, the effective temperature in the Nariai limit is greater than that of the extremely hot black hole.
This indicates that the steering ${\cal G}^{A \to B}_3$ recovers its initial value $\ln[\cosh(2s)] $ for the extremely hot black hole in SdS spacetime. This similar conclusion has been used \cite{QWE70}.

\section{ Conclusions  \label{GSCDGE}}
The gravitational effect of the black hole and an expanding universe
on Gaussian quantum steering and its asymmetry in the Schwarzschild-de Sitter black hole spacetime has been investigated. This spacetime is endowed with black hole event horizon (BEH) as well as  cosmological event horizon (CEH). By placing a thermally opaque membrane in between the two horizons, two independent thermal-equilibrium regions are constructed. Our model involves two modes:
the  mode $A$ observed by Alice located at BEH; the mode $B$ observed
by Bob located at CEH. We have found that quantum steering between Alice and Bob reduces monotonically with the Hawking temperature of the black hole and may cause sudden death, but changes non-monotonically with the Hawking temperature of the expanding universe.
This means that the Hawking effect of the expanding universe is not always harmful to Gaussian quantum steering.
We have also studied the asymmetry of quantum steering and found that Alice who locates in the higher temperature of BEH has stronger steerability than Bob who locates in the lower temperature of CEH. This result is interesting and nonintuitive, because the thermal noise introduced by the Hawking temperature destroys quantum steering. We have presented the conditions for two-way, one-way, and no-way steering, and found that the maximal steering asymmetry
always occurs at the transition from one-way to two-way steering. Finally, we have demonstrated the relationship between steering and entanglement. We have found that Gaussian quantum steering suffers from a sudden birth with the increase of the initial squeezing, indicating that quantum steering becomes feasible only when entanglement reaches some levels.

On the other hand, we have studied Gaussian quantum steering from the scenario of effective temperature which is associated with both the Hawking temperature of the black hole and the Hawking temperature of the expanding universe. In this scenario, particle creation with the effective temperature occurs in a single region $r_H<r<r_C$ and not in causally disconnected spacetime wedges. In contrast to the effective temperature, particle creation near the BEH and CEH with the Hawking temperature  occurs  in causally disconnected spacetime wedges. We have found that the Gaussian quantum steering reduces monotonically with the effective temperature. Actually, Gaussian quantum steering in this scenario increases monotonically with the Hawking temperature of the  black hole no matter how hot the black hole is. This is because  the effective temperature increases with the growth of the mass  of the black hole. In other words, reducing the Hawking temperature of the black hole actually increases the effective temperature. Therefore,  the effective temperature and Hawking temperature of the black hole have different effects on Gaussian quantum steering due to their differences.

\begin{acknowledgments}
This work is supported by the National Natural
Science Foundation of China (Grant Nos. 12205133, 12175060 and 12075050), LJKQZ20222315 and JYTMS20231051.	
\end{acknowledgments}

$\textbf{Data Availability Statement}$

This manuscript has no associated data.

$\textbf{Conflict of interest}$

The authors declare no conflicts of interest.


\end{document}